\documentclass[useAMS,usenatbib]{mn2e}
\usepackage{epsfig}
\usepackage[totalwidth=480pt, totalheight=680pt]{geometry} 

\def\beq{\begin{equation}}
\def\eeq{\end{equation}}

\def\msol{\,\mathrm{M}_\odot}

\def\a{(a)}
\def\b{(b)}
\def\C{(c)}
\def\d{(d)}

\title[Nuclear Cluster formation through Globular Cluster decay and merging]
{Self-consistent simulations of Nuclear Cluster formation through Globular Cluster orbital decay
and merging}
\author[R. Capuzzo-Dolcetta, P. Miocchi]{R. Capuzzo-Dolcetta\thanks{E-mail:
Roberto.Capuzzodolcetta@uniroma1.it} P. Miocchi\thanks{E-mail:
miocchi@uniroma1.it}\\
Dipartimento di Fisica, ``Sapienza'' Universit\'a di Roma , 
P.le A. Moro, 5, Roma I-00185, Italy.}
\begin{document}

\date{Accepted ????. Received ????; in original form ????}

\pagerange{\pageref{firstpage}--\pageref{lastpage}} \pubyear{????}

\maketitle

\label{firstpage}

\begin{abstract}
We present results of fully self-consistent $N$-body simulations of the
motion of four globular clusters moving in the inner region of their 
parent galaxy. With regard to previous simplified simulations, we
confirm merging and formation of an almost steady nuclear cluster, 
in a slightly shorter time. The projected surface density profile shows
strong similarity to that of resolved galactic nuclei. This similarity
reflects also in the velocity dispersion profile which exhibits a
central colder component as observed in many nucleated galaxies.  
\end{abstract}

\defcitealias{cdm08}{CM08}

\section{Introduction}
The motion of massive globular clusters (GCs) in a galaxy is determined by both the 
smooth general potential and the graininess of the stellar field. This latter,
fluctuating, component acts as a decelerating mechanism: the so called dynamical friction
(hereafter DFR).
The role of DFR, which dissipates cluster orbital energy and angular momentum,
has been found to be crucial in determining the time
evolution of globular cluster orbits in a galaxy, especially for orbits plunging
in the inner, high density galactic regions \citep[e.g.][]{cdv05}.
The decay time may, indeed, be short enough for massive clusters to limit their
motion to the inner galactic regions. However, any general astrophysical
consideration regarding dynamical friction cannot be based just on 
approximated evaluations based on Chandrasekhar-like formulas, even in their
generalizations apt to treat axisymmetric or triaxial cases (\citealt*{pcv}; \citealt{cd93};
\citealt{cdv05}).

Several effects are indeed neglected by the analytical approach. For example,
it is not considered the change in magnitude of the DFR due to the increase of the
spatial extension of the cluster caused by the formation, and subsequent `expansion', of its
tidal tails, and, moreover, how the DFR acts on the different parts of the cluster
(e.g., it could be stronger on its core than on the tails).
Recent numerical experiments have tried to shed light on this aspect 
finding that stars stripped from the cluster by the field, but still close enough
to the system, continue to contribute to the mass of the decelerating system \citep{fell07},
and that, in general, the real DFR effect can be stronger than that estimated by the usual
Chandrasekhar formula \citep[probably because of the further friction due to tidal effects,
see][]{funato}.
Another unexplored question is how the gravitational feedback on the very inner part of the
galaxy influences the DFR strength; this may be important when, during the final
stages of orbital decay, the GC orbit gets very near to the centre so to enclose a galactic
mass comparable with that of the cluster itself.

Clarifying the role of the above-mentioned dynamical effects
is important also in the attempt to understand the mechanisms leading to the formation of
Nuclear Clusters (NCs). In fact, it can help in discriminating the various scenarios
proposed, especially in supporting the validity of multiple merging of smaller 
sub-systems as a formation mechanism \citep[see, e.g.][]{oh00,fell02,bekki04,cdm08}.
At this regard, we cite the recent observational evidence of the existence of a very young
and massive star cluster in NGC 2139, located $2\arcsec$ \textit{offset} from the kinematical centre
of the host galaxy, suggesting a formation that was independent and non coeval with that of the
galactic bulge and, moreover, the observed environment is such that
the system can decay to the centre in a time so short to keep intact its structure and become
what is normally called NC (\citealt{andersen08}, see also \citealt{boker02,walcher06}).  

Reliable indications would follow by straightforward, direct $N$-body
integrations, which can be enormously time consuming when treating
self-consistently the motion of a GC in a dense galactic environment.
In this Letter we present the preliminary results of a fully 
self-consistent $N$-body simulation concerning the close interaction 
of a sample of four massive GCs in the central region of a galaxy.
Both the clusters and the galaxy are represented by mutually interacting particles,
thus including self-consistently both DFR and tidal interactions. 
This study can give depeer insight into the decay and merging of GCs in 
galactic nuclear regions, a scenario first tackled by semi-analitical 
approaches \citep{trem,cd93} and then pursued by one full $N$-body experiment 
though with a resolution much lower than that presented here (\citealt{oh00}, 
who used $N=500$ `particles' for each GC and $N\sim 10^4$ to represent the galaxy).
Simulations presented here are useful to test and validate previous recent
results \citep[][hereafter CM08]{cdm08} obtained in a simplified scheme in which
the clusters are $N$-body objects moving in a fixed analytical galactic 
potential and subjected to a dynamical friction braking given analitically by the
\citet{pcv} formula. 

\section{Simulations}
Our $N$-body simulation has been carried out employing a parallel tree-code with a
leap-frog time integrator using individual and variable time steps \citep{cdm02}.
The simulation is set in a way similar to that described in
\citetalias{cdm08}, but for the role of the external galactic field which was, there,
represented as a triaxial, analytical potential while, here, it is self-consistently
($N$-body) represented starting from initial conditions sampling a given equilibrium
density profile. We resume briefly the main characteristics of our modelization.
We choose to follow four massive GCs as templates of a multi-merger among stellar systems
decayed, conserving their individuality, in the inner galaxy region.
A quick orbital decay induced by DFR is due to initial large values for the total
mass of the clusters, whose resistance to external tidal disturbance was guaranteed
by their sufficient initial compactness. The GC sample corresponds to the four densest clusters
dealt with in \citetalias{cdm08} and, moreover,
every GC is composed by $N=2.5 \times 10^5$ equal mass stars.
See Tab.~\ref{tab1} for the initial values of their structural parameters.

The stellar bulge where GCs move is here represented as another $N$-body system,
with $N_\mathrm{b}=512,000$ particles, sampling a Plummer model such to give the
bodies initial positions and velocities according to a Montecarlo representation of
its equilibrium distribution function ($\propto [-E]^{7/2}$ with $E$ being 
the particle total energy per unit mass).
The Plummer density profile
\beq
\rho_\mathrm{b}(r)=\rho_{\mathrm{b}0}\left[1+(r/r_\mathrm{bc})^2\right]^{-5/2},
\eeq
where $\rho_{\mathrm{b}0}=3\sqrt{8}M_\mathrm{b}/4\pi r_\mathrm{bc}^3$ is the central density,
has two scale parameters, the core mass ($M_\mathrm{b}$)
and the core radius ($r_\mathrm{bc}$) which are used, in the following, as mass and length units,
respectively.
Consequently, time, velocity and density will be measured in units of the galactic crossing time
$t_\mathrm{b}=(r_\mathrm{bc}^3/GM_\mathrm{b})^{1/2}$, of
$v_\mathrm{b}=(GM_\mathrm{b}/r_\mathrm{bc})^{1/2}$ and of the central density
$\rho_{\mathrm{b}0}$, respectively. 
Due to the infinite extension of the Plummer sphere, it has to be truncated at a distance
$r_\mathrm{cut}$ large enough to guarantee the stability of the sampling $N$-body representation.
This is done by the choice of $r_\mathrm{cut}= 12.2 r_\mathrm{bc}$ so as to contain 99\% of
the total mass.
In order to have nearly the same galactic mean gravitational field of that in the \citetalias{cdm08}
model, we set the $M_\mathrm{b}$ and $r_\mathrm{bc}$ values as equal to those of the
corresponding parameters
of the galactic density distribution used in \citetalias{cdm08}, even though this gives a
slightly larger $\rho_{\mathrm{b}0}$.   
The four clusters [\a, \b, \C\ and \d] start their evolution from the same initial conditions
chosen in \citetalias{cdm08}. As clusters reference centres, we chose their centre-of-densities
(CDs) as defined in \citet{casertano}.
Finally, note that the simulation results are scale-invariant: any quantity can be
re-scaled by fixing arbitrarily the values of $M_\mathrm{b}$ and $r_\mathrm{bc}$ in physical units. 

\begin{figure*}
\includegraphics[width=8.5cm]{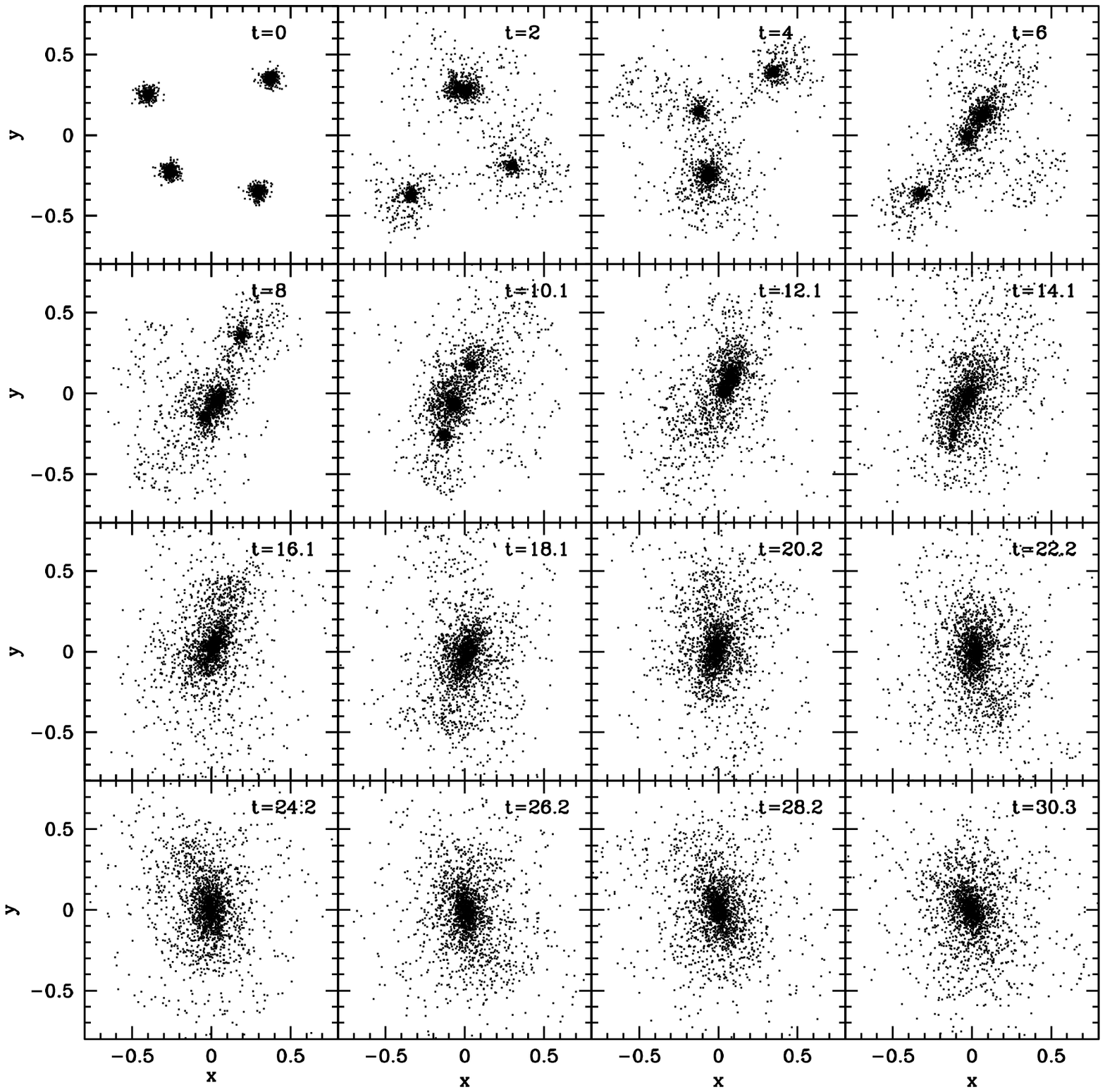}
\hfill
\includegraphics[width=8.5cm]{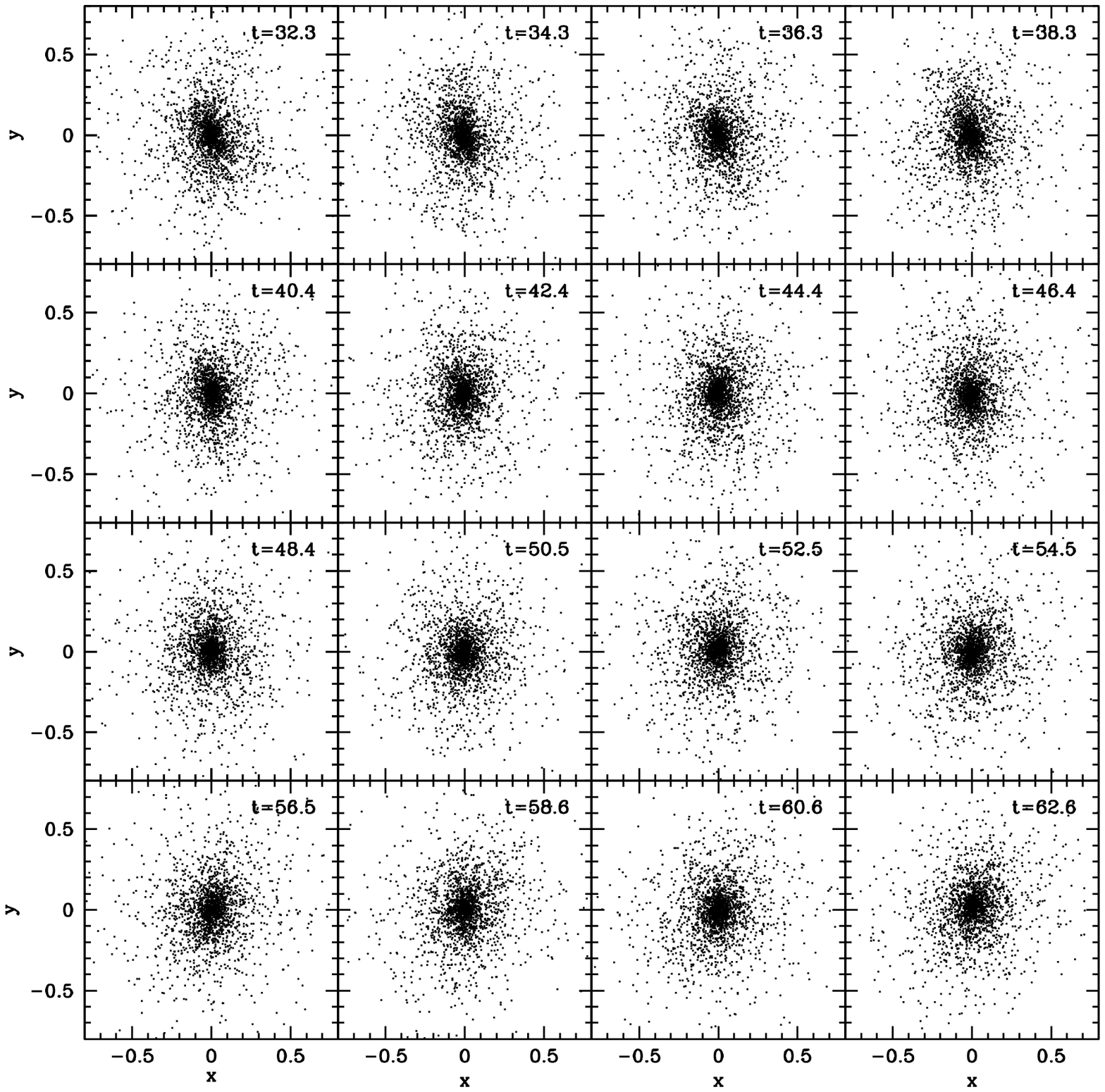}
\caption{
Time sequence of the four GCs merger. Field stars are not shown.
\label{snap}}
\end{figure*}

\begin{figure}
\includegraphics[width=8.5cm]{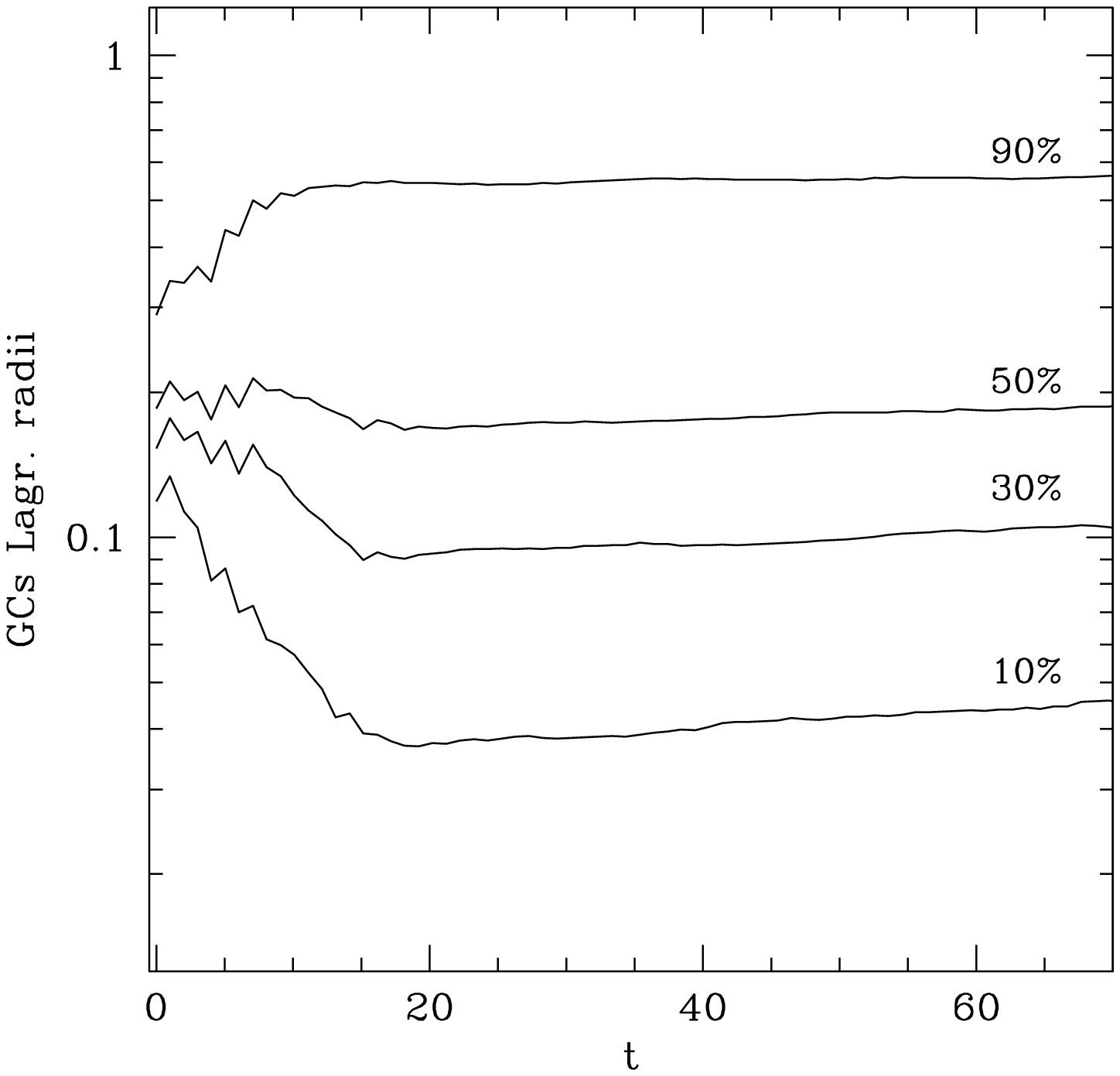}
\vfill
\includegraphics[width=8.5cm]{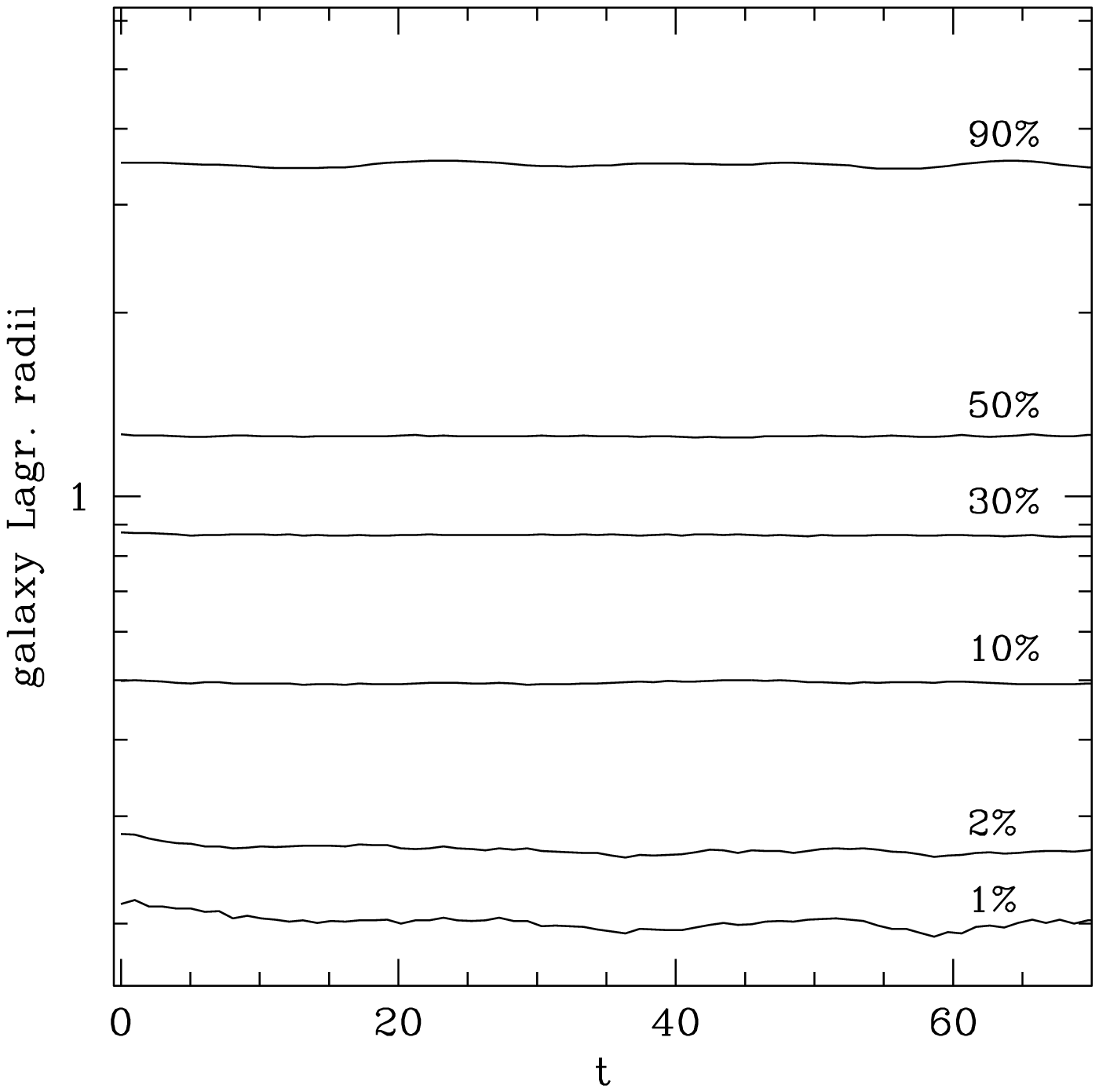}
\caption{Evolution of the Lagrangian radii of the
four GCs as a whole (upper panel) and of the galaxy (lower panel)
averaged over a time window $=3t_\mathrm{b}$. They correspond to
the radii of the sphere centered at the system CD and enclosing a given fraction
of the total mass (as indicated).
\label{rlag}}
\end{figure}

\begin{figure}
\includegraphics[width=8.5cm]{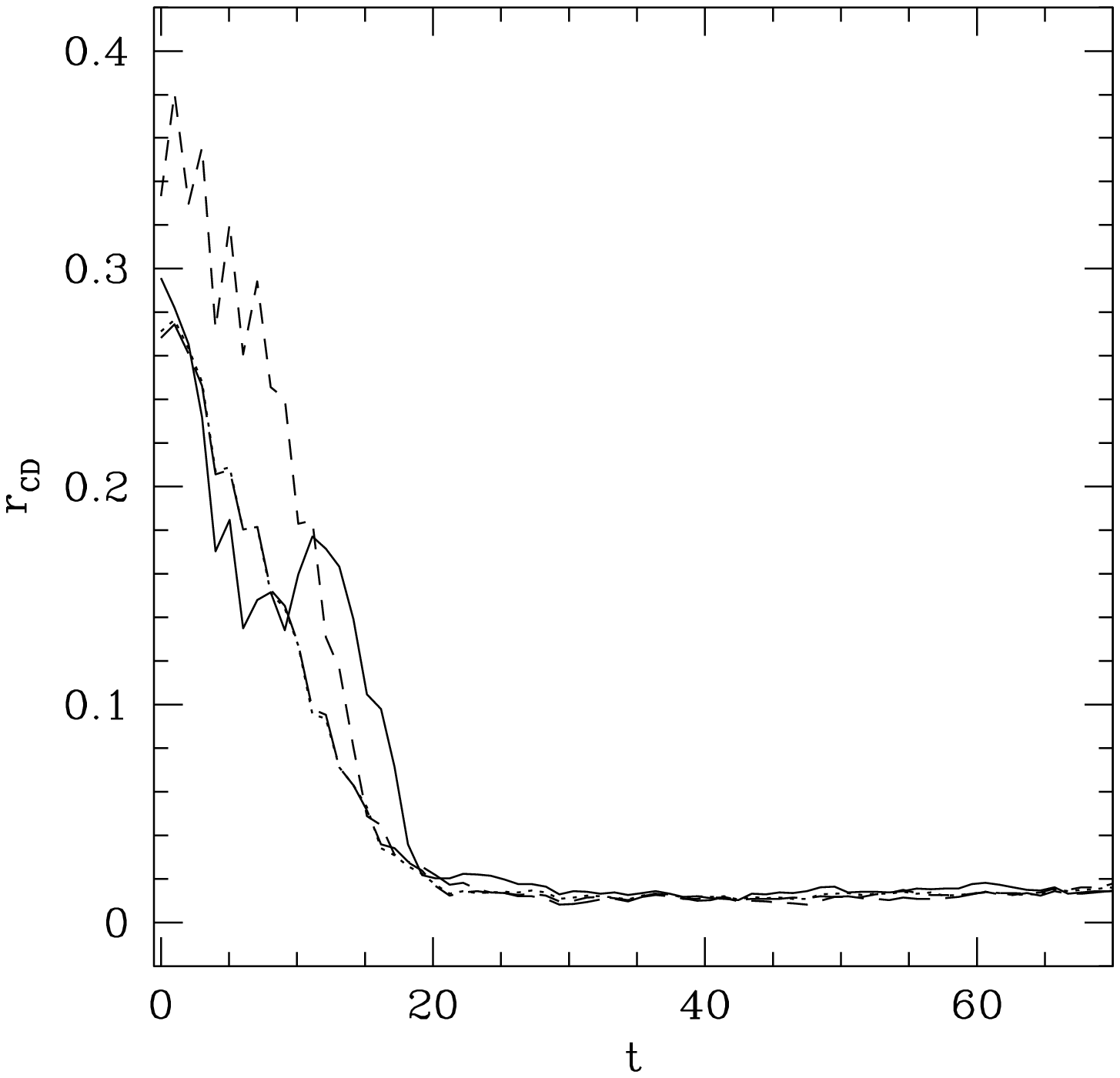}
\vfill
\includegraphics[width=8.5cm]{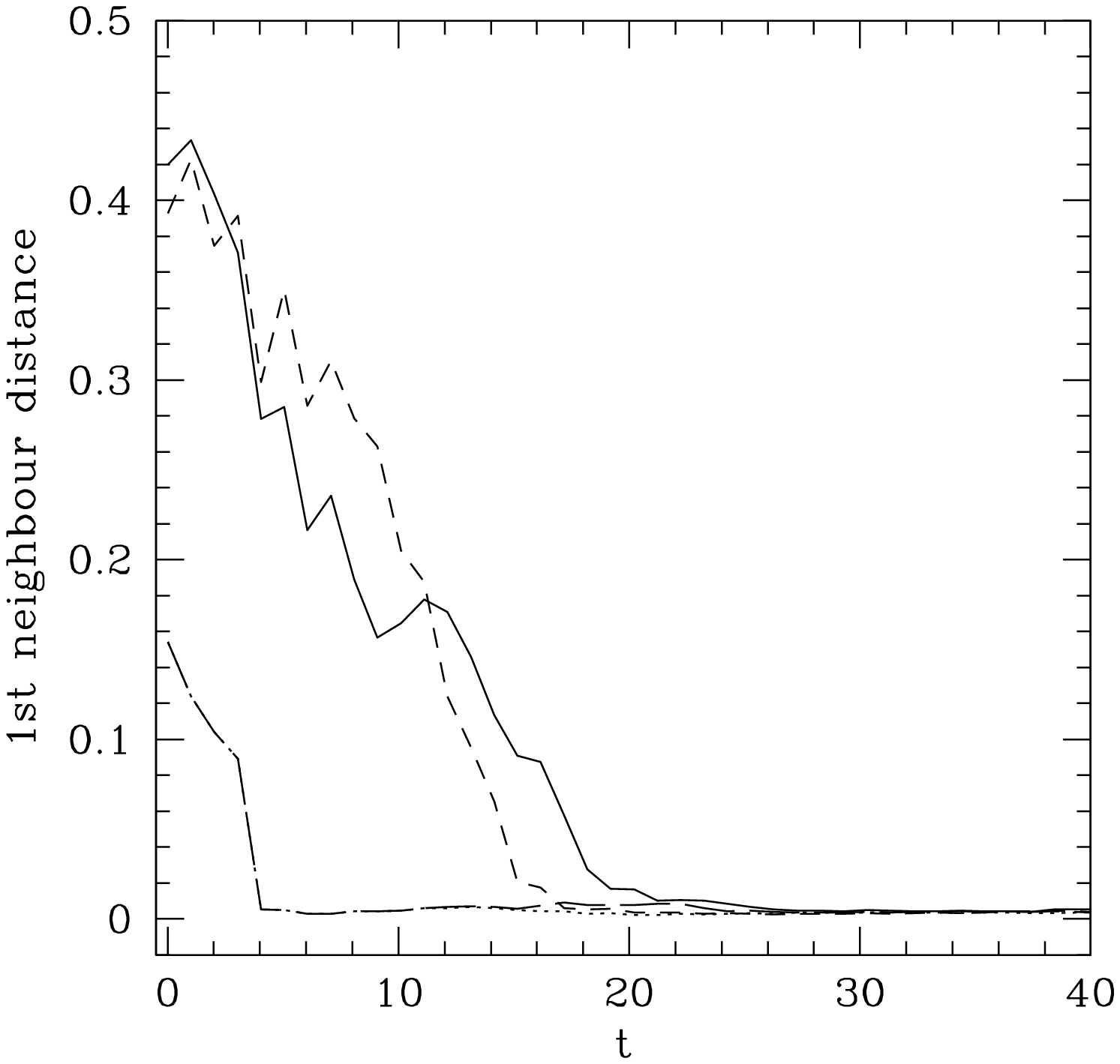}
\caption{Upper panel: time evolution of the distance of the cluster CD from the galactic
centre ($r_\mathrm{CD}$).
Lower panel: time behaviour of the distance between each cluster CD and its first-neighbour.
All the behaviours have been averaged over a time window $=3t_\mathrm{b}$.
Solid lines refer to cluster \a; dotted: \b; short-dashed: \C; long-dashed: \d.
\label{rcd}}
\end{figure}

\begin{table*}
 \centering
 \begin{minipage}{122mm}
\caption{Parameters list for the initial cluster models,
expressed in galactic units.
Reported are: the GC mass ($M$), the limiting radius ($r_\mathrm{t}$),
the King radius ($r_\mathrm{c}$), the King
concentration parameter ($c$),  the half-mass radius
($r_\mathrm{h}$), the central density ($\rho_0$),
the half-mass crossing time, $t_\mathrm{ch}\equiv [r_\mathrm{h}^3/(GM)]^{1/2}$, and
the King velocity parameter ($\sigma_\mathrm{K}$).
\label{tab1}}
  \begin{tabular}{@{}p{0.8cm}cccccccc@{}}
\hline

cluster & $M$ & $r_\mathrm{t}$ & $r_\mathrm{c}$ &  $c$ & $r_\mathrm{h}$ & $\rho_0$
 & $t_\mathrm{ch}$ & $\sigma_\mathrm{K}$\\
\hline

{\bf \a}&
$6.7\times 10^{-3}$&
$0.16$&
$1.1\times 10^{-2}$&
$1.2$&
$2.3\times 10^{-2}$&
$510$&
$3.9\times 10^{-2}$&
$0.24$\\
{\bf \b}&
$7.6\times 10^{-3}$&
$0.16$&
$1.5\times 10^{-2}$&
$1.0$&
$2.4\times 10^{-2}$&
$270$&
$4.2\times 10^{-2}$&
$0.24$\\
{\bf \C}&
$8.1\times 10^{-3}$&
$0.14$&
$1.4\times 10^{-2}$&
$0.99$&
$2.2\times 10^{-2}$&
$400$&
$3.4\times 10^{-2}$&
$0.27$\\
{\bf \d}&
$6.3\times 10^{-3}$&
$0.14$&
$1.8\times 10^{-2}$&
$0.89$&
$2.4\times 10^{-2}$&
$160$&
$4.9\times 10^{-2}$&
$0.23$\\
\hline
\end{tabular}
\end{minipage}
\end{table*}


\section{Results}
The rapidity of the merger is evident in Fig.~\ref{snap}, as well as in
the upper panel of Fig.~\ref{rcd}: the merger process is completed at about
$t_\mathrm{m}\simeq 17t_\mathrm{b}$, when the Lagrangian radii of the four GCs, 
seen as a whole system, flatten to a quasi-stable configuration 
(see Fig.~\ref{rlag}, upper panel). In physical units:
\beq
t_\mathrm{m}\simeq 0.5 \left(\frac{r_\mathrm{bc}}{100 \textrm{ pc}}\right)^{3/2}\left(\frac{M_\mathrm{b}}
{10^9 \msol}\right)^{-1/2}
\textrm{Myr}.
\eeq
Of course this `merging' time depends much on the orbital initial conditions 
of the progenitor clusters and measures the time of the merger since the time 
in which the GCs are already confined within the galactic core region. 
Nevertheless, given that DFR has been convincingly shown to be an efficient mechanism
to drag GCs in the very inner regions of a galaxy within a time much 
shorter than a Hubble time (see \citealt{cd93}, \citealt{cdv05} and the 
discussion in \citealt{mio06}), this result supports the hypothesis
stating that a GCs `sedimentation' can take place at the galactic centre well 
within the galaxy life-time.

It is interesting noting (see the upper panel of Fig.~\ref{rcd}), that the CD of the merger
remnant oscillates within $0.02r_\mathrm{bc}$ from the galactic CD.
If, e.g., $r_\mathrm{bc}\sim 200$ pc this displacement corresponds
to $\sim 4$ pc, i.e. a separation that would be hard to be resolved
(corresponding to $\sim 0.05 \arcsec$ at the Virgo cluster distance), so the NC would
appear substantially centered. 
The off-centered NC position is due to the gravitational feedback between
the very central part of the galaxy and the clusters, which naturally derives from 
the self-consistent nature of the model. Such an interaction reflects into the `perturbed'
behaviour of the 1\% Lagrangian radius (below $\sim 0.2 r_\mathrm{bc}$) of the galaxy 
that is evident from Fig.~\ref{rlag} (lower panel).

The lower panel of Fig.~\ref{rcd}, indicates that a first merging event between 2 clusters 
[\b\ and \d] occurs rather early (at $t<5$), when their centres are still 
$\sim 0.2 r_\mathrm{bc}$ far from the galactic centre.
The fact that such a binary coalescence takes place \textit{before} the complete
orbital decay, suggests that these events can occur even offset from
the galactic centre, in spite of the strong tidal field. This `pre-decay' coalescence 
was also found in the low-resolution self-consistent simulation by \citet{oh00} 
(their model 1b), though in their case this phenomenon takes place
much farther from the galactic centre.

\begin{figure}
\includegraphics[width=8.5cm]{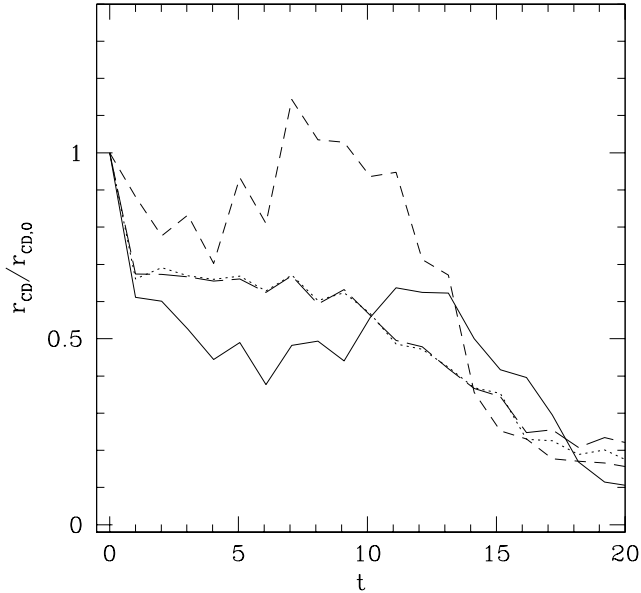}
\caption{Time evolution of the ratio between the distance of the cluster 
CD from the galactic centre and the same quantity evaluated in the \citetalias{cdm08} simulation
($r_{\mathrm{CD},0}$), averaged over a time window $=3t_\mathrm{b}$.
Line symbols are the same as in Fig.~\ref{rcd}.
\label{rcdconfr}}
\end{figure}

By comparing the merger time with that of the \citetalias{cdm08} simulation, it is found that
here the orbital decay is $\sim 1.6$ times faster than that given by the use (in \citetalias{cdm08})
of the generalized \citep{pcv} Chandrasekhar formula


This means that the DFR efficiency is greatly enhanced in the self-consistent model. 
Indeed, in Fig.~\ref{rcdconfr} we see that, at least for $t<20$, the clusters CDs 
are always \textit{closer} to the centre of the galaxy respect to the case of the 
\citetalias{cdm08} simulation. For later times the comparison is no longer 
reliable, because of the very small values of $r_\mathrm{CD}$.

\begin{figure}
\includegraphics[width=8.5cm]{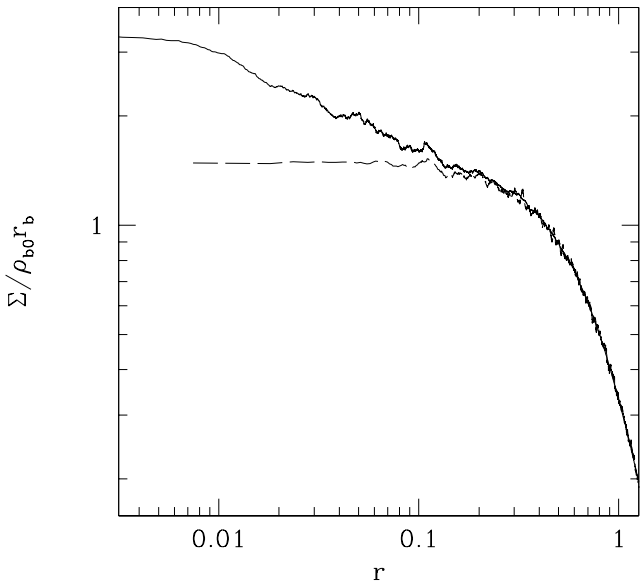}
\caption{Projected surface density profile of the whole system (galaxy plus NC) in the final
configuration (solid line) and, for the sake of comparison, of the galaxy only (long dashed line).
\label{surf}}
\end{figure}

\begin{figure}
\includegraphics[width=8.5cm]{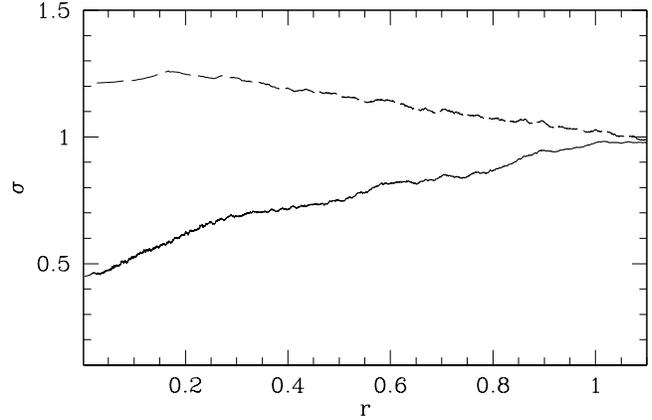}
\caption{Projected radial behaviour of the velocity dispersion in the
last system configuration. Line symbols are as in Fig.~\ref{surf}.
\label{sigma}}
\end{figure}

In Fig.~\ref{surf} the total surface density profile is plotted for the last configuration
and the typical appearance of a nucleated galaxy central profile comes clearly out.
As an example, the similarity with the VCC 1871 profile in the Virgo cluster \citep{cote06}
is particularly evident. The central `over-density' associated
to the NC presence is, however, less pronounced with respect to that obtained in
\citetalias{cdm08} and this is due to the lower concentration of the NC in comparison
with the last configuration in \citetalias{cdm08} case. Perhaps, the lower NC central
density may be a consequence of the too large collisional effect generated by the interaction
of the clusters with the `particles' representing the very inner region of the galaxy that,
because of the limited resolution, are probably too massive. Only
higher resolution simulations of the very last stages of the orbital decay and merging
would allow to establish the actual role of this effect (Capuzzo-Dolcetta \& Miocchi, in preparation).

An important result concerns the peculiar behaviour of the global velocity dispersion
profile. Confirming the finding in \citetalias{cdm08}, the velocity dispersion
of the whole system (galaxy plus NC) shows a clear decrease towards the centre
(see Fig.~\ref{sigma}).
As discussed in \citetalias{cdm08}, this is a clear sign of the presence of two kinematically
distinct systems that are relaxed into a different dynamical status. Such a peculiar feature
has been actually found in most of the Virgo cluster nucleated dwarf ellipticals observed by
\citet*[][see their Fig.~5]{geha}. The decreasing trend to the centre of the velocity dispersion
is also consistent with the solution of the Jeans
equations for a sample of NCs observed in late-type spirals \citep{walcher05}.
Finally, either a flattening
or a slight central decrease are found and discussed in \citet{oh00}, too.


\section{Summary}
Massive globular clusters suffer of dynamical friction braking such to be limited
to move in the inner part of the host galaxy. The interaction of decayed globular
clusters among themselves and with the external tidal field may lead to subsequent
merger events, whose modes are not easily understood without detailed $N$-body simulations.
In a previous paper \citepalias{cdm08} we followed numerically the interaction and merging
of four globular clusters each represented as an $N=250,000$ stars system moving in an
analitical triaxial potential, taking dynamical friction into account by means of
the \citet{pcv} formula.

The present paper has the aim to check the \citetalias{cdm08} results
via a full self-consistent $N$-body modelization, i.e. sampling also the galactic environment
by a large number ($=512,000$) of interacting particles.
All the main features of the merging event and of the `super cluster' remnant found
by \citetalias{cdm08} are qualitatively confirmed by this, more detailed, simulation,
but for a stronger orbital decay (two times faster) due to collective effects
and for a less centrally concentrated merger product.
This latter result, however, needs to be checked by higher resolution simulations
which are also necessary to study reliably the long-term evolution and stability of
the formed NC.

\section*{Acknowledgments}
The simulation was conducted at the CINECA supercomputing centre, under the
INAF-CINECA agreement (grant \textit{cne0in07}).

\onecolumn

\bsp

\label{lastpage}


\begin{thebibliography}{99}
\bibitem[\protect\citeauthoryear{Andersen et al.}{2008}]{andersen08}
Andersen D.R., Walcher C.J., B\"oker T., Ho L.C., van der Marel R.P., Rix H.-W., Shields J.C.,
2008, submitted to ApJ.
\bibitem[\protect\citeauthoryear{Bekki et al.}{2004}]{bekki04}  Bekki K., Couch W.J., Drinkwater M.J.,
Shioya Y., 2004, ApJ, 610, L13
\bibitem[\protect\citeauthoryear{B\"oker et al.}{2002}]{boker02} B\"oker T., Laine S.,
van der Marel R.P., Sarzi M., Rix H.-W., Ho L.C., Shields J.C., 2002, AJ, 123, 1389
\bibitem[\protect\citeauthoryear{Capuzzo-Dolcetta}{1993}]{cd93} Capuzzo-Dolcetta R., 1993, ApJ, 415, 616
\bibitem[\protect\citeauthoryear{Capuzzo-Dolcetta \& Miocchi}{2008}]{cdm08} Capuzzo-Dolcetta R.,
Miocchi P., 2008, accepted on ApJ (astro-ph/0801.1072) \citepalias{cdm08}
\bibitem[\protect\citeauthoryear{Capuzzo-Dolcetta \& Vicari}{2005}]{cdv05}Capuzzo-Dolcetta R.,
Vicari A. 2005, MNRAS, 356, 899
\bibitem[\protect\citeauthoryear{Casertano \& Hut}{1985}]{casertano}Casertano S., Hut P., 1985, ApJ,
298, 80
\bibitem[\protect\citeauthoryear{C\^ot\'e et al.}{2006}]{cote06} C\^ot\'e P. et al., 2006, ApJS, 165, 57
\bibitem[\protect\citeauthoryear{Fellhauer et al.}{2002}]{fell02}
Fellhauer M., Baumgardt H., Kroupa P., Spurzem R., 2002, Cel. Mech. Dyn. Astron., 82, 113
\bibitem[\protect\citeauthoryear{Fellhauer \& Lin}{2007}]{fell07}Fellhauer M., Lin, D.N.C., 2007,
MNRAS, 375, 604
\bibitem[\protect\citeauthoryear{Fujii et al.}{2007}]{funato}Fujii M., Iwasawa M., Funato Y., Makino J.,
2007, submitted to ApJ (astro-ph/0708.3719)
\bibitem[\protect\citeauthoryear{Geha, Guhathakurta \& van der Marel}{Geha et al.}{2002}]{geha}
Geha M., Guhathakurta P., van der Marel R.P., 2002, AJ, 124, 3073
\bibitem[\protect\citeauthoryear{Miocchi \& Capuzzo-Dolcetta}{2002}]{cdm02} Miocchi P., Capuzzo-Dolcetta R.,
2002, A\&A, 382, 758
\bibitem[\protect\citeauthoryear{Miocchi et al.}{2006}]{mio06} Miocchi P., Capuzzo-Dolcetta R.,
Di Matteo P., Vicari A., 2006, ApJ, 644, 940
\bibitem[\protect\citeauthoryear{Oh \& Lin}{2000}]{oh00}Oh K.S., Lin D.N.C., 2000, ApJ, 543, 620
\bibitem[\protect\citeauthoryear{Pesce, Capuzzo-Dolcetta \& Vietri}{Pesce et al.}{1992}]{pcv}
Pesce E., Capuzzo-Dolcetta R., Vietri M., 1992, MNRAS, 254, 466
\bibitem[\protect\citeauthoryear{Tremaine, Ostriker \& Spitzer}{Tremaine et al.}{1975}]
{trem} Tremaine S., Ostriker J.P., Spitzer L. Jr., 1975, ApJ, 196, 407
\bibitem[\protect\citeauthoryear{Walcher et al.}{2005}]{walcher05} Walcher C. J. et al.,
2005, ApJ, 618, 237
\bibitem[\protect\citeauthoryear{Walcher et al.}{2006}]
{walcher06} Walcher C.J., B\"oker T., Charlot S., Ho L.C., Rix H.-W., Rossa J., Shields J.C.,
van der Marel R.P., 2006, ApJ, 649, 692
\end{thebibliography}
\end{document}